 \UseRawInputEncoding 
\documentclass[
reprint,
 superscriptaddress,
 amsmath,amssymb,
 aps,
prl,
floatfix,
]{revtex4-2}

\usepackage[version=4]{mhchem}
\usepackage{graphicx}
\usepackage{dcolumn}
\usepackage{xcolor}
\usepackage{bm}
\usepackage{physics}

\usepackage[english]{babel}

\begin{document}

\preprint{APS/123-QED}


\title{Demonstration of the ODMR activity of the telecom range ClV center in SiC: a wavefunction theory analysis}

\author{Zsolt Benedek}
\email{benedek.zsolt@ttk.elte.hu}
\author{Oscar Bulancea-Lindvall}%
\affiliation{Department of Physics of Complex Systems, E\"otv\"os Lor\'{a}nd University, Budapest, Hungary}
\affiliation{MTA–ELTE Lend\"{u}let "Momentum" NewQubit Research Group, Budapest, Hungary}
\affiliation{Department of Physics, Chemistry and Biology (IFM), Linköping University}

\author{Joel Davidsson}
\affiliation{Department of Physics, Chemistry and Biology (IFM), Linköping University}
\author{Viktor Iv\'{a}dy}
\affiliation{Department of Physics of Complex Systems, E\"otv\"os Lor\'{a}nd University, Budapest, Hungary}
\affiliation{MTA–ELTE Lend\"{u}let "Momentum" NewQubit Research Group, Budapest, Hungary}
\author{Igor Abrikosov}
\affiliation{Department of Physics, Chemistry and Biology (IFM), Linköping University}

\date{\today}

\begin{abstract}
Recently, density functional theory-based high-throughput screening of point defects in 4H-SiC revealed the positively charged chlorine-vacancy (\ce{ClV}) defect to be a promising quantum bit candidate emitting at telecom wavelengths, with an electronic structure analogous to the well-known \ce{NV} center in diamond. Furthermore, recent infrared photoluminescence (PL) measurements on chlorine-implanted 4H-SiC have revealed new PL lines associated with the ClV defect. While the defect possesses a high-spin ground state, there is a lack of evidence of optically detected magnetic resonance (ODMR), a key ingredient for optical spin initialization and readout. 
In this Letter, we employ a multireference wavefunction-based quantum chemistry method, specifically, second-order perturbation theory (NEVPT2) on top of a defect-localized many-body wavefunction (CASSCF), to explore the many-body electronic structure of the ClV center. We estimate photoluminescence, internal conversion, and intersystem crossing rates to investigate the possibility of spin polarization and ODMR activity.
Our findings establish the \ce{ClV} center in 4H-SiC as an optically addressable spin qubit with fiber optics compatibility in the technologically mature 4H-SiC host material, enabling the development of large-scale quantum networks.
\end{abstract}

\maketitle



\noindent Color centers in wide-bandgap semiconductors have been studied for over two decades due to their potential in emerging quantum technologies, including spin-photon interfaces~\cite{xiongHighthroughputIdentificationSpinphoton2023,baiIdentificationLuminescentDefects2025,castellettoSiliconCarbideColor2020}, single-photon emitters~\cite{sipahigilIntegratedDiamondNanophotonics2016a,cilibrizziUltranarrowInhomogeneousSpectral2023a}, spin qubits~\cite{bradleyTenQubitSolidStateSpin2019,andersonFivesecondCoherenceSingle2022a,zhangMaterialPlatformsDefect2020a}, and quantum sensors~\cite{jiangQuantumSensingRadiofrequency2023,luoHighsensitivitySiliconCarbide2023,tarasenkoSpinOpticalProperties2018,vacancy_relaxometry}. 
Considerable attention has been paid to the negatively charged nitrogen vacancy (NV) center in diamond, demonstrating great versatility in applications thanks to its remarkable quantum properties, such as high-temperature coherence and strong visible  emission~\cite{herbschlebUltralongCoherenceTimes2019,linTemperaturedependentCoherenceProperties2021}.
The success of the NV center inspired the search for NV-like defects in other host materials. For example, several recent works have focused on color centers in 4H silicon carbide (4H-SiC) with similar characteristics to harness the fabrication advantages of the technologically mature SiC host~\cite{lukin4HsiliconcarbideoninsulatorIntegratedQuantum2020,yiSiliconCarbideIntegrated2022}.

\begin{figure*}[t]
    \centering
    \includegraphics[width=\linewidth]{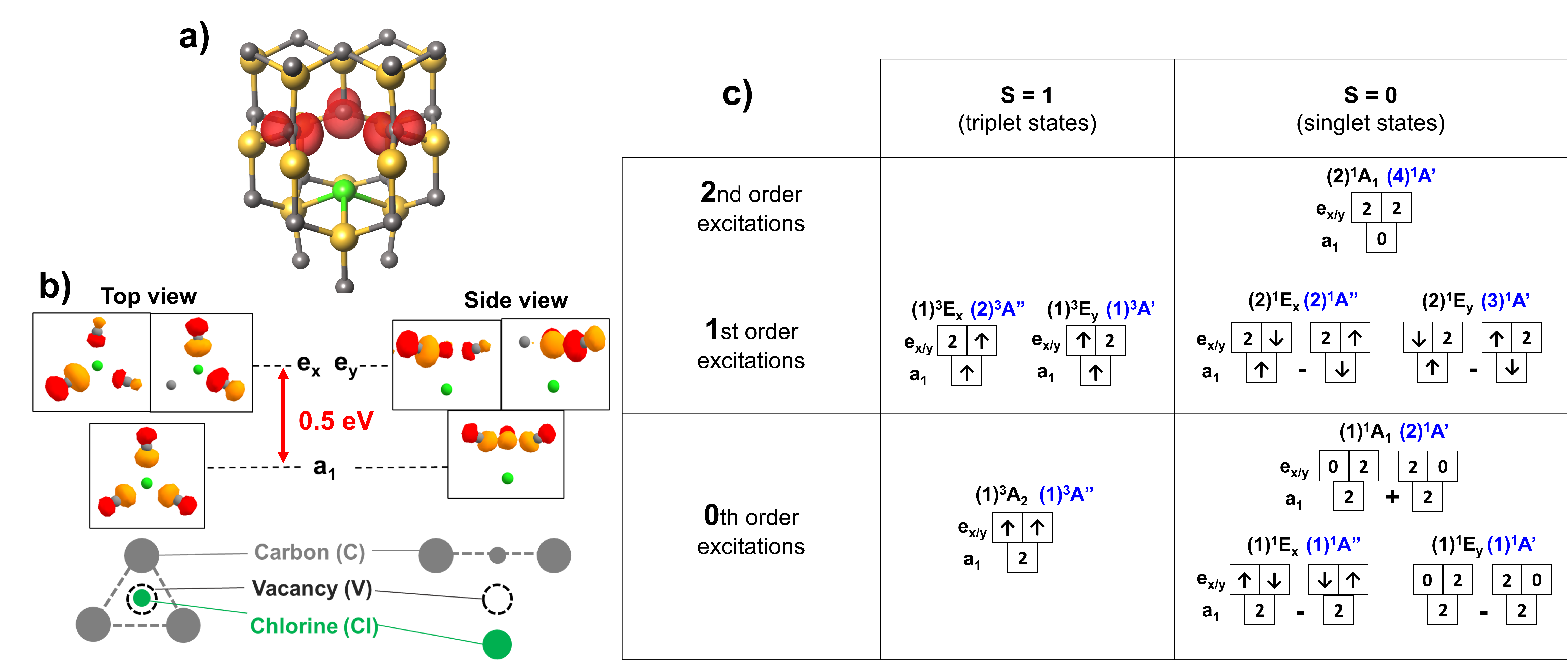}
    \caption{a) Crystal structure and spin density of the \ce{ClV} defect in 4H-SiC, in the ground state $kk$ configuration b) Shape of frontier molecular orbitals. c) Conceivable occupation patterns of $e_x, e_y$ and $a_1$ orbitals according to group theory analysis. Black and blue signs indicate irreducible representations at $C_{3v}$ and $C_{s}$ symmetry, respectively.}
    \label{fig:orbs+confs}
\end{figure*}

Automated computer simulation tools have accelerated the pace of defect search by performing density functional theory (DFT) on a large number of defect structures.  As such, the high-throughput workflow implemented in the ADAQ code~\cite{adaq} revealed the positively charged \ce{ClV} defect center in 4H-SiC, see Fig.~\ref{fig:orbs+confs}a, as a near-C-band telecom emitter~\cite{Oscar}, making it the most favorable spin qubit candidate in terms of compatibility with quantum communication networks utilizing modern fiber optic technologies at minimal signal loss.
The triplet ($S = 1$) ground spin state, the similarity to the \ce{NV} center in electronic structure, and the predicted long coherence time~\cite{Oscar} of the \ce{ClV} center could also enable its application as a qubit and quantum sensor~\cite{khrapkoQuasiSingleModeFiber2024}. 

Importantly, very recent experiments have reported on the observation of telecom S-band, O-band, and C-band spectral lines in 4H-SiC after chlorine implantation~\cite{anisimovEngineeringChlorinebasedEmitters2025}. The four previously unobserved ZPL lines have been tentatively assigned to the four possible configurations of the \ce{ClV} center in SiC, based on the circumstances of creation and the overall good agreement of the theoretical and experimental emission energies \cite{anisimovEngineeringChlorinebasedEmitters2025}.

Despite these promising recent reports,  several questions remain open on the properties and the applicability of the \ce{ClV} center. It is vital for potential spin qubit applications that its spin state can be optically polarized and read out via spin-selective photoluminescence. When the electronic structure of the defect enables these, simultaneous application of optical excitation and a microwave drive resonant with a given spin state transition induces variation of the PL intensity, often referred to as optically detected magnetic resonance, observed and predicted for many defects in the field~\cite{suterOpticalDetectionMagnetic2020,thieringTheoryOpticalSpinpolarization2018,dongSpinPolarizationIntersystem2019a,bianTheoryOpticalSpinpolarization2025}.


Computationally demonstrating that a defect is ODMR active requires a comprehensive characterization of the system's optical cycle, including the optically excited spin-1 many-body state, the metastable spin-0 many-body shelving state(s), and radiative and non-radiative transitions among them. The electronic states often include open-shell states, which are not accurately modeled in DFT due to their multireference character~\cite{galiInitioTheoryNitrogenvacancy2019b,chenMulticonfigurationalNatureElectron2023b}. The ${}^1A_1$ shelving state of the NV center is one such state notorious for its inaccuracy in DFT and other single-determinant methods \cite{chenMulticonfigurationalNatureElectron2023b,jinVibrationallyResolvedOptical2022a}. Due to the analogy between the \ce{NV} center in diamond and the \ce{ClV} center in 4H-SiC, similar hurdles can be expected in the theoretical analysis of the latter. To describe such states, higher-order multireferential computational techniques must be employed. A cost-efficient example is CASSCF-NEVPT2 (i.e., n-electron valence state perturbation theory~\cite{nevpt2,Angeli-2001a,Angeli2007} on top of a complete active space self-consistent field~\cite{Siegbahn_1980,Roos_1980,Siegbahn_1981} wavefunction), which has been demonstrated to accurately describe the excited states of \ce{NV-} in diamond~\cite{benedekAccurateConvergentEnergetics2024,Luu-2025}.

In this Letter, we employ a large-scale 151-atom SiC cluster-model implementation of the CASSCF-NEVPT2 methodology, as available in current quantum chemical codes~\cite{Guo_2021,Kollmar_2021,dlpno-nevpt2}, to achieve a complete and accurate description of the ClV center's many-body electronic structure. We utilize this model to address the open questions regarding the excitation dynamics of the defect and discuss the implications of the predicted transition rates. We conclude that the ClV center can be spin polarized by optical excitation, giving rise to ODMR signals when microwave irradiation is additionally applied. These results establish the ClV center as a telecom wavelength atomic-scale spin qubit.




To this end, we carry out wavefunction theory-level computations using the ORCA program~\cite{neese2022software}, version 6.0.1.  
To build a proper model of ClV center-containing 4H-SiC, two hydrogen-terminated cluster models were prepared, among which the smaller (\ce{Si12C28H64Cl+}) was used for method testing purposes, while the larger (\ce{Si_{56}C_{95}H_{156}Cl^+}) was used for production calculations. According to our previous works~\cite{benedekAccurateConvergentEnergetics2024,Luu-2025}, the latter model is sufficiently large to closely reproduce the properties of a defect in an infinite lattice. In 4H-SiC, the \ce{ClV} defect can be constructed in four different configurations, i.e. $kk$, $hh$, $kh$, and $hk$~\cite{Oscar}. For computational efficiency and to convey the main message of this Letter, we only discuss the electronic structure of the $kk$ configuration in detail. The quantitative description of other configurations will be presented elsewhere. We demonstrate, however, that the electronic structures of the $hh$, $kh$, and $hk$  configurations are qualitatively the same, see section S1.7 of the supplementary information (SI)~\cite{supplemental-info}, containing Refs.~\cite{Roos1987,Olsen-2011,Angeli-2001,Chilkuri2021,Chilkuri2021b,Sayfutyarova2017,cc-pVDZ,NEESE2009,Weigend2006,Weigend2002,Roemelt_2013,doi:Neese1998,Neese_2005,Schatz1993-kb,PhysRevB.104.045303,Smart2021,PhysRevB.100.081407,PhysRevB.90.075202,PhysRevB.110.184302,Illg2016-bg,RevModPhys.73.515,PhysRevB.90.075202,ivadyFirstPrinciplesCalculation2018b}.


For the CASSCF calculations, we selected a 3-orbital active space, containing the in-gap $a_1$ and $e$ orbitals localized to the inner carbons, see Fig.~\ref{fig:orbs+confs}b. These orbitals are occupied by four electrons, which is possible in several different configurations, see Fig.~\ref{fig:orbs+confs}c, for a group theory analysis. State-specific CASSCF(4e,3o) calculations were performed to obtain the equilibrium geometry of the lowest-lying triplet and singlet electronic states on a wavefunction theory level. On the equilibrium geometries, single-point state-average [3S+3T]-CASSCF(4e,3o)-NEVPT2 calculations were carried out to evaluate energies and properties, where [3S+3T] indicates three triplet and three singlet states, involved in the averaging scheme with equal weights. Further details on method testing and sample input files are available in section S1 of the SI~\cite{supplemental-info}.



The main characteristics of the electronic states of the \ce{ClV} defect can be described by the occupation pattern of three frontier molecular orbitals ($a_1, e_x, e_y$, shown in Fig.~\ref{fig:orbs+confs}b), formed as linear combinations of the dangling sp3 orbitals of carbon atoms adjacent to the vacancy. We note that the fourth dangling sp3 orbital is strongly localized to the chlorine atom and is buried deeply in the valence band. 

\begin{figure*}[t]
    \centering
    \includegraphics[width=\textwidth]{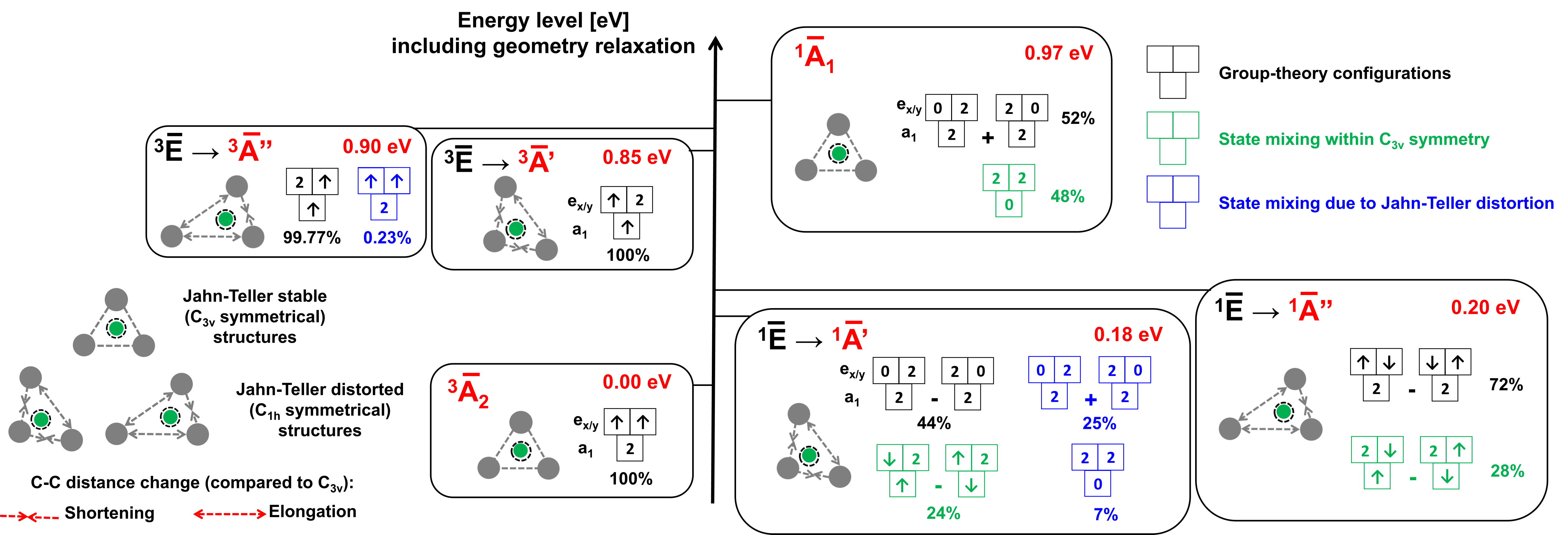}
    \caption{Many-body composition (on the basis of canonical $e_{x/y}$ and $a_1$ orbitals) and energy level of the lowest-lying electronic states of \ce{ClV} in 4H-SiC, computed at CASSCF-NEVPT2 level of theory. Black, green, and blue color orbital occupation patterns indicate group-theory configurations, state mixing within $C_{3v}$ symmetry, and state mixing related to Jahn-Teller distortion, respectively.}
    \label{fig:relaxed}
\end{figure*}

The conceivable electronic configurations predicted by group theory in $C_{3v}$ and $C_{1h}$ symmetry are depicted in Fig.~\ref{fig:orbs+confs}c. These pure states are mixed to form the many-body eigenstates at the CASSCF level,  
which is visualized in Fig.~\ref{fig:relaxed}. In the following, the six lowest-lying eigenstates, which are expected to be relevant for qubit applications, will be referred to according the nomenclature shown in Fig.~\ref{fig:relaxed}, i.e. ${}^3\bar{A}_2$, ${}^3\bar{E}_x $ (${}^3\bar{A}'$),
${}^3\bar{E}_y $ ($ {}^3\bar{A}''$),${}^1\bar{E}_x $ ($ {}^1\bar{A}'$), ${}^1\bar{E}_y $ (${}^1\bar{A}''$) in  $C_{3v}$ ($C_{1h}$), and ${}^1\bar{A}_1$, where the overline emphasizes the distinction of the mixed many body states from the unmixed group-theory states. 



The mixing of group-theory states, as provided by percentages in Fig.~\ref{fig:relaxed}, can be attributed to two main factors. Firstly, the high-lying and low-lying states of equivalent $C_{3v}$ irreducible representations are already intensively mixed at high-symmetry geometries: that is, $^1\bar{E}$ and $^1\bar{A}_1$ states borrow Slater determinants from multiple group-theory configurations, $(1)^1E$ and $(2)^1E$ or $(1)^1A_1$ and $(2)^1A_1$, respectively.  Green occupation diagrams show this type of mixing in Fig.~\ref{fig:relaxed}. Secondly, the symmetry reduction to $C_{1h}$ due to Jahn-Teller distortion of the $E$ states, enables additional combinations within $A'$ and $A''$ symmetry blocks. The electron configurations appearing due to the latter effect are shown in blue in Fig.~\ref{fig:relaxed}.    
Jahn-Teller distortion of the excited state, in particular ${}^3E_x\rightarrow {}^3\bar{A}''$, introduces the character of the ground triplet state ($ 0.23\%\ {}^3\bar{A}_2$). This JT-induced state mixing enables a non-radiative internal conversion (IC) between the optically excited triplet and the ground state, which competes with the desired telecom-band photoemission.


In the case of singlets, the proportion of state mixing at $C_{3v}$ symmetry is approximately 70\%-30\% between $(1){}^1E$ and $(2){}^1E$, which appear in ${}^1E\rightarrow {}^1\bar{A}''$ and ${}^1E\rightarrow {}^1\bar{A}'$, and a striking near-equal contribution between $(1){}^1A_1$ and $(2){}^1A_1$, which form the many-body state of ${}^1\bar{A}_1$. Upon symmetry reduction by Jahn-Teller distortion, ${}^1E\rightarrow {}^1\bar{A}'$ also becomes mixed with $(1){}^1A_1$ and $(2){}^1A_1$. Conspicuously, these new contributions to ${}^1E\rightarrow {}^1\bar{A}'$ (32\% in total) are far more significant than the nearly undetectable mixing of $A''$ type triplet states, which can be traced back to the identical Slater determinants in $(1){}^1E_y$ and $(1){}^1A_1$, see Fig.~\ref{fig:orbs+confs}c.


Altogether, the composition of singlet and triplet states is analogous to the many-body picture of \ce{NV} center~\cite{benedekAccurateConvergentEnergetics2024}, albeit the degree of multireference character of singlets is considerably higher, which has an important effect on the transition rates.


The positions of energy level of the studied electronic states on relaxed geometries, relative to the ground state of $^3\bar{A}_2$, are shown in Fig.~\ref{fig:relaxed}. We note that a detailed discussion of these relaxations, including $C_{3v}$ symmetry conserving relaxation and symmetry breaking Jahn-Teller relaxation, can be found in the SI~\cite{supplemental-info}, section S2. Based on our wavefunction theory calculation, the lowest energy triplet excited state ${}^3\bar{A}'$ is located at 0.85~eV, giving rise to a $\sim1460$~nm ZPL emission for the $kk$ configuration of the ClV center. This value confirms S-band telecom emission and corresponds well to our previously reported DFT calculations~\cite{Oscar}.  
Furthermore, the small difference ($\approx$50 meV) between the two excited energy levels indicates dynamic Jahn-Teller behavior, analogously to the corresponding ${}^3E$ state of \ce{NV} center in diamond~\cite{thiering_ab_2017}.

Our multireference calculations can be used to position the singlet states. The lowest-lying singlets, ${}^1\bar{E} \rightarrow {}^1\bar{A}'$ and ${}^1\bar{E} \rightarrow {}^1\bar{A}''$ are located 0.18 and 0.20 eV above the ground triplet, respectively. Similar to the triplet excited state, dynamic Jahn-Teller character was found at the CASSCF-NEVPT2 level of theory. The third singlet state, $^1\bar{A}_1$ lies at 0.97 eV, which is slightly (0.07-0.12 eV) above the level of ${}^3\bar{E}$ state. We emphasize that this energy difference is comparable to the expected error margin of our wavefunction theory calculations.



\begin{figure}[h!]
    \centering
    \includegraphics[width=0.7\linewidth]{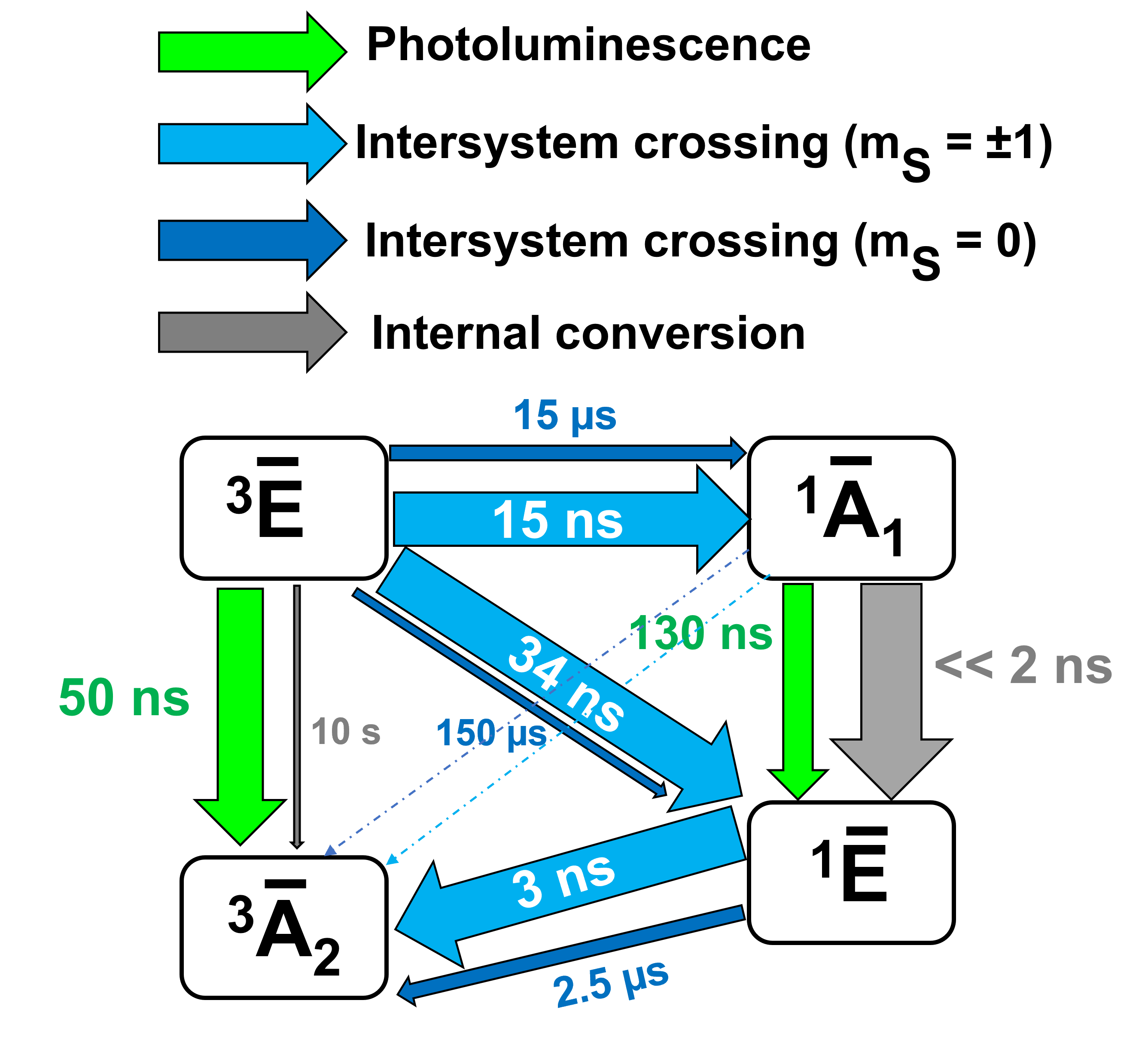}
    \caption{Calculated transition rates (given as half-lives) between the lowest-lying electronic states of \ce{ClV}@4H-SiC. }
    \label{fig:rates}
\end{figure}

Having all equilibrium geometries and energy levels at hand, we fitted 1D effective phonons to all possible transitions among the six identified low-lying electronic states. Applying Fermi's golden rule by the equations described in SI~\cite{supplemental-info}, section S1.5, we were able to calculate all photoluminescence, intersystem crossing, and internal conversion rates. These results are visualized in Fig.~\ref{fig:rates}, with more details on numerical results, such as transition matrix elements, displacements, and effective phonon energies, in Table S4 of SI~\cite{supplemental-info}. 

Firstly, let us discuss the spin conserving relaxations (${}^3\bar{E} \rightarrow {}^3\bar{A}_2$ for triplets, ${}^1\bar{A}_1 \rightarrow {}^1\bar{E}$ for singlets). In the triplet sector, this process is radiation-dominated ($\tau_{\rm{PL}}$~=~50~ns). This can be traced back to the high transition dipole moment,$\approx$7 Debyes, facilitating PL as well as the negligible triplet wavefunction mixing, 0.23\% ${}^3A_2$ in ${}^3A''$, see Fig.~\ref{fig:relaxed}, which drives a slow nonradiative internal conversion with a half-life in the range of seconds. The singlet-to-singlet conversion, on the other hand, comes with a remarkably fast IC ($\tau_{\rm IC}$~=~2~ns). The vast deviation from the triplet IC rate can be explained by the intensive state mixing through the pseudo Jahn-Teller effect between ${}^1\bar{A}_1$ and ${}^1\bar{A}'$, as indicated by the large weight of Jahn-Teller distortion-related singlet configurations in Fig. 2. Here, the competing photoluminescence, which occurs at a comparable rate to triplet-triplet radiative transition, is about two orders of magnitude slower ($\tau_{\rm PL}$ = 130 ns). We further note here that Fermi's golden rule assumes weak coupling between the initial and final states, which might not hold for the closely related ${}^1\bar{A}_1$ and ${}^1\bar{A}'$ species; therefore, in our opinion, the obtained nonradiative lifetime of 2 ns should be considered as an upper limit. 


Apart from photon emissions and the ${}^1\bar{A}_1 \rightarrow {}^1\bar{E}$ internal conversion, intersystem crossings (ISC) connecting the singlet-triplet spin sectors can also occur at a nanosecond timescale. The ISC processes of ${}^3\bar{E} \rightarrow {}^1\bar{A}_1$, ${}^3\bar{E} \rightarrow {}^1\bar{E}$ and ${}^1\bar{E} \rightarrow {}^3\bar{A}_2$, which are driven by strong spin-orbit coupling matrix elements (10-40 GHz), proceed with half-lives of 15~ns, 34~ns, and 3~ns respectively. 
However, these fast rates only apply to the $m_S = \pm 1$ channel in the triplet state; the conversion to/from the $m_S = 0$ sublevel was calculated to be about three orders of magnitude slower in all three cases,
owing primarily to the reduced parallel spin-orbit coupling matrix elements compared to the perpendicular $m_S = \pm 1$ case.  



For bright emission from the ClV defect, the ${}^3\bar{E} \rightarrow {}^3\bar{A}_2$ transition should not be suppressed by alternative nonradiative relaxation routes. Based on our kinetic model, this requirement is fulfilled, i.e., the direct IC in the triplet sector has negligible rate, while ISC towards either ${}^1\bar{A}_1$ or ${}^1\bar{E}$ has at most similar rate to PL, even if we consider the kinetically favored ($m_S = \pm 1$) channel. This means that the photoemission is expected to be clearly detectable even from the $m_S = \pm 1$ sublevel of $^3\bar{E}$, albeit ISC might reduce the quantum yield to some extent. Furthermore, at $m_S = 0$ sublevel, PL is clearly the dominant relaxation mechanism. In other words, $m_S = 0$ can be viewed as the "bright" state, in contrast to the "dark" $m_S = \pm 1$, which is more apt to relaxation via ISC. 

Another important characteristic is the relative intensity of the ZPL in the emission spectrum, i.e., the Debye-Waller factor (DWF). Previous DFT work reported on a 1-3\% DWF (depending on the configuration~\cite{Oscar}), calculated for a single phonon connecting the geometry of the Jahn-Teller distorted excited state and that of the ground state. We hereby confirm this DWF range by CASSCF-NEVPT2 computations, see SI~\cite{supplemental-info}, Section S4, for details. Thus, the emission of phonons during PL also significantly reduces the quantum yield towards the desired narrow ZPL sign, but only to a similar extent to the case of \ce{NV} center in diamond. However, we note that the DWF can be potentially increased by Purcell enhancement in SiC crystal cavities~\cite{Crook2020}.

As for the singlet (${}^1\bar{A}_1 \rightarrow {}^1\bar{E}$) emission, our calculation suggests a similarly low DWF factor (0.7 \%). This PL is expected to be barely detectable, as IC is dominant, and even the remaining PL consists mainly of a broad phonon sideband.



As ISCs predicted for all triplet spin sublevels (0, $\pm$1) with significant rates (ns-$\mu$s  timescale), we concluded from our results that an optical spin polarization cycle similar to that of the NV center in diamond is possible. However, even though the relative rate of $m_S = \pm 1$ and $m_S = 0$ channels is different, it cannot be unequivocally determined at the present level of theory which sublevel of ${}^3\bar{A}_2$ is populated.
Furthermore, as mentioned previously, the $m_S = \pm 1$ and $m_S = 0$ sublevels of ${}^3\bar{A}_2$ can be viewed as the dark and bright states, respectively, owing to the vastly different ISC rates. This enables optical spin readout and also gives rise to an ODMR signal.




Having shown evidence for the \ce{ClV} in 4H-SiC to be ODMR active, its ODMR signal could be one of the most relevant characteristics for identification as it carries information on the unique fine and hyperfine structure of the ground state of the \ce{ClV} center. To simulate the ODMR signal, we first calculate the ZFS of the ClV center. 
Even though D values on DFT level have been previously reported~\cite{Oscar}, the calculations carried out in this work allow us to provide revised results. Indeed, the previous unrestricted DFT computations suffer from spin contamination issues, which can be resolved by (i) CASSCF-NEVPT2 level zero-field splitting calculations, which also operate over restricted open-shell orbitals in quantum chemical programs, (ii) restricted open-shell (RO) Kohn-Sham formalism, (iii) spin de-contamination corrections~\cite{biktagirov_spin_2020} on the previous unrestricted results. We recall that due to the single-reference character of $^3\bar{A}_2$, DFT level is expected to be sufficiently accurate for this property.
In the spin de-contamination procedure of the unrestricted DFT, we used the Vienna Ab-initio Simulation Package~\cite{kresseEfficientIterativeSchemes1996a,kresseInitioMoleculardynamicsSimulation1994a} which implements Kohn-Sham DFT based on the projector-augmented wave method~\cite{kresseUltrasoftPseudopotentialsProjector1999b,blochlProjectorAugmentedwaveMethod1994a}, on 576-atom supercells with PBE using a 2x2x2 $k$-point grid.

For the cluster model with $kk$ defect configuration, we obtain 1.34 GHz and 1.02 GHz at CASSCF-NEVPT2 and RO-DFT level, respectively. Similarly, the supercell model of Ref.~\onlinecite{Oscar} gives $D=1.15$ GHz, taking into account spin de-contamination, details on the revision of supercell DFT calculations can be found in section S5 of SI~\cite{supplemental-info}. While the methodologies differ on the scale of ~300 MHz, they fairly agree on a corrected range for the \ce{ClV} ZFS parameters.




We present here a qualitative investigation into possible ODMR signatures for the \ce{ClV} in 4H-SiC. With the spin parameters for the ClV center hyperfine coupling to the component Cl isotope from Ref.~\onlinecite{Oscar}, which is dominantly a $I = 3/2$ nuclear spin, and updated ZFS tensors obtained in this work, an effective spin model of the ClV center electron spin and Cl nuclear spin can be constructed and diagonalized to reveal its possible microwave-driven transitions, see SI~\cite{supplemental-info}, section S6 for details.  


\begin{figure}
    \centering
    \includegraphics[width=0.95\linewidth, clip, trim=0.55cm 0.0cm 1.25cm 1.00cm]{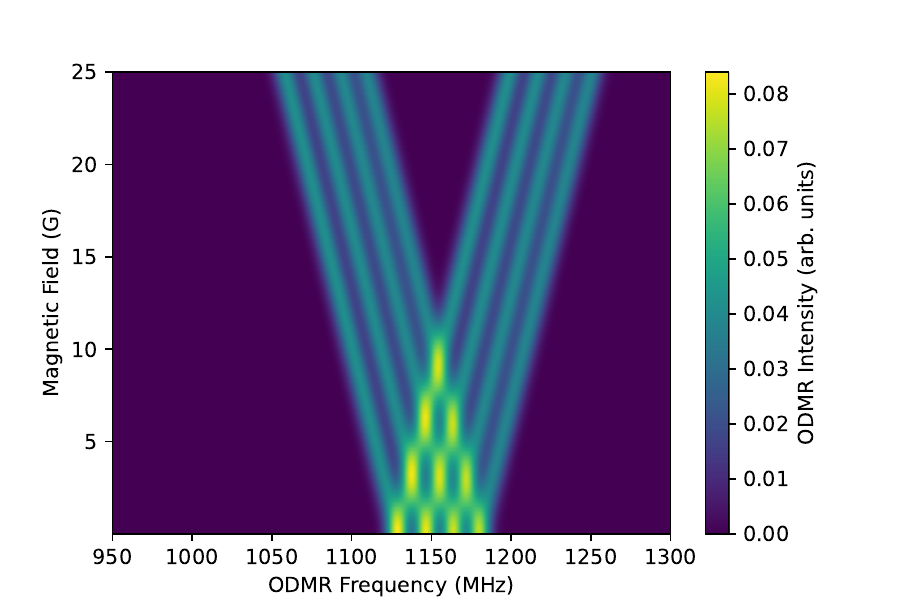}
    \caption{Microwave drive matrix element amplitudes of the ClV${}^+$ and Cl spin system for the $kk$ configuration, incorporating ${}^{35}$Cl. A Gaussian broadening term is applied in the frequency domain, using a standard deviation of 5 MHz.}
    \label{fig:kk_ODMR}
\end{figure}

Figure~\ref{fig:kk_ODMR} shows our ODMR reconstruction for the $kk$ configuration, in the given range showing the electron transition between $m_S = \pm 1$ and $m_S = 0$ levels of the \ce{ClV} electron spin. We note that the Cl isotope spin quartet will contribute with a splitting of the signature origin by a multiple of its hyperfine, at ca.\ 15 MHz. If the experimental frequency or field resolution is insufficient to resolve this hyperfine, the $kk$ and $hh$ clearly exhibiting this splitting will likely appear as fairly broad spectral signatures in the ODMR. The reader is referred to Section S6 of the SI~\cite{supplemental-info} for a more detailed discussion.


In conclusion, we investigated the lowest-energy electronic states of the \ce{ClV} defect in 4H-SiC using multireference wavefunction theory. Our calculations strongly suggest that the defect is ODMR active and is therefore suitable for quantum computing and quantum sensing applications. In experiments, the defect can be identified based on its telecom-band ZPL (0.8-0.9 eV) and its ODMR sign corresponding to D = 1.0-1.3 GHz zero-field splitting. 


\emph{Data availability} --- The main data supporting the findings of this study are available within the paper and its Supplementary Information. Further numerical data are available from the authors upon reasonable request.

\begin{acknowledgments}

\emph{Acknowledgments} --- We thank Danial Shafizadeh, Ivan G. Ivanov, and Tien Son Nguyen for the fruitful discussions.  Z.B. acknowledges the financial support of János Bolyai Research Fellowship of the Hungarian Academy of Sciences.
J.D. acknowledges support from the Swedish Research Council (VR) Grant No. 2022-00276 and 2020-05402. We acknowledge support from the Knut and Alice Wallenberg Foundation (Grant No. 2018.0071). I.A.A. is a Wallenberg Scholar (Grant No KAW 2023.0309).
This research was supported by the National Research, Development, and Innovation Office of Hungary within the Quantum Information National Laboratory of Hungary (Grant No. 2022-2.1.1-NL-2022-00004) and within grant FK 145395.
This project is funded by the European Commission within Horizon Europe projects (Grant Nos.\ 101156088 and 101129663). The computations were enabled by resources provided by the National Academic Infrastructure for Supercomputing in Sweden (NAISS), partially funded by the Swedish Research Council through grant agreement no. 2022-06725. We acknowledge KIF\"U for awarding us access to computational resources in Hungary.

\end{acknowledgments}

\bibliography{apssamp}

\end{document}


\maketitle
\date


\tableofcontents
\newpage

\section{Theoretical section (more details on the applied WFT methodology)}

\subsection{Construction of cluster models}

\begin{figure}[h]
    \centering
    \includegraphics[width=0.5\linewidth]{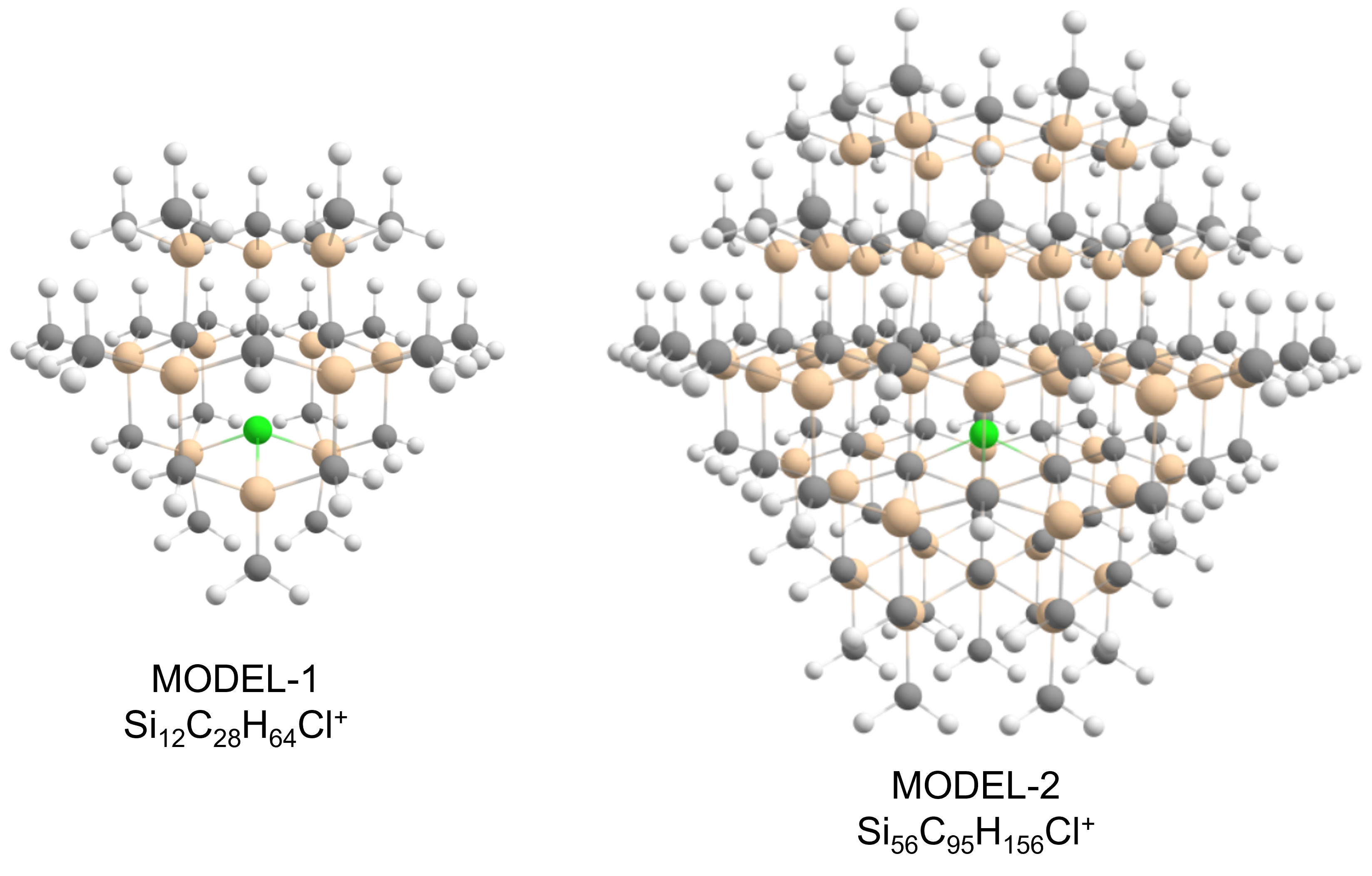}
    \caption{Visualization of the molecular model sizes used for the present study. In the main text, the results for MODEL-2 will be discussed, while test calculations using MODEL-1 are detailed in the SI. Color code of atoms: green - Cl, brown - Si, grey - C, white - H.}
    \label{fig:models}
\end{figure}

For the time being, wavefunction-based quantum chemical calculations are mostly available for molecular models. Therefore, we modeled the structure of \ce{ClV+}@4H-SiC defects by hydrogen terminated clusters, which were prepared as follows. In the first step, an Si atom was replaced by a vacancy in a large 4H-SiC cell, the atomic positions in which were determined according to experiments. Then, we selected Si and C atoms in the order of the distance from the vacancy. Atoms within 3 chemical bond distance relative to the vacancy, along with the terminating H atoms and after a carbon$\rightarrow$chlorine replacement, give our MODEL-1 (Fig.~\ref{fig:models}). This small cluster, which contains one layer of C and one layer of Si atoms around the innermost shell (i.e. \ce{C_3Cl}), is only suitable for qualitative investigations (which were performed for the purpose of method development). Upon the expansion of the model by another layer of Si and another layer of C atoms, we arrive to MODEL-2, the size of which is - according to our previous works~\cite{benedekAccurateConvergentEnergetics2024,Luu-2025} - sufficiently large to provide quantitatively correct results.

\subsection{Wavefunction theory}

\subsubsection{Static correlation (CASSCF)}

We calculated the many-body wavefunctions for the electronic states of \ce{ClV+} at state-average (SA) complete active space self-consistent field (CASSCF)~\cite{Roos1987,Olsen-2011,casscf} level of theory, with NEVPT2 perturbation~\cite{Angeli-2001,Angeli2007,Guo_2021,Kollmar_2021} accounting for dynamic correlation.

Within CASSCF, we take all possible electron configuration patterns within a selected molecular orbital subspace into account. During the SA-CASSCF calculation, a two-step cycle of full configuration interaction (FCI) solution and orbital optimization (i.e. mixing active and external orbitals to minimize the average energy of the considered electronic states) is repeated until convergence. For the SA-CASSCF procedure, one needs to choose (i) the number of electronic states to be involved, (ii) the active orbitals, (iii) the number of electrons that occupy the latter orbitals in various patterns. Our considerations on these input parameters are detailed below. 

As for the number of electronic states, we selected 3 triplet and 3 singlet roots. Namely, based on our initial density functional theory (DFT) results~\cite{Oscar}, we expected \ce{ClV+}@4H-SiC to form analogous electronic structures to \ce{NV-}@diamond, where the spin polarization and spin-selective readout is realized via 3 triplet and 3 singlet low-lying states (${}^3A_2, {}^3E_{x/y}, {}^1E_{x/y}, {}^1A_1$).

Next, let us consider the number of relevant orbitals. Upon the formation of a \ce{ClV+} vacancy in the 4H-SiC lattice, 4 dangling orbitals are formed as linear combinations of partially occupied $sp^3$ orbitals of C and Cl atoms surrounding the vacancy. Nevertheless, it is not evident that this "chemically motivated" active space is the best suited for CASSCF-based methods. For example, the "dangling" $sp^3$ orbital of chlorine is deeply buried in the valence band, and is therefore expected to be inert in CASSCF calculations. Furthermore, non-dangling orbitals might also contribute to static correlation significantly. 

For these reasons, we performed approximate CASSCF calculations on the full valence space of the atoms surrounding the \ce{ClV+} defect (i.e. 2s and 2p shells of the innermost three C and three Si atoms, as well as the Cl atom), in order to rationally select the active space. We utilized the recently introduced iterative configuration expansion (ICE) method~\cite{Chilkuri2021,Chilkuri2021b} as an approximate CI solver, and ran the calculation on the atomic valence active space (AVAS)~\cite{Sayfutyarova2017}, prepared for the aforementioned atoms and atomic orbitals. This ICE-SCF calculation (48 electrons, 28 orbitals, 3 triplet and 3 singlet electronic states) revealed that only the $a_1$, $e_{x}$ and $e_y$ orbitals possess significant fractional occupation numbers (1.97$>$occ.$>$0.03) in natural orbital basis. This result implies the reliability of a three-orbital active space.

On these three carbon-centered orbitals, 4 valence electrons of the innermost atoms are distributed; these are the "free" electrons that are not involved in covalent C-Si and Cl-Si bonds or in the chlorine localized  $a_1^{\rm(low)}$ orbital (1 free electron from each of the three C atoms adjacent to the vacancy; 2 free electrons from Cl; 1 electron is subtracted due to the net positive charge). Altogether, we arrive at 4-electron (4e) 3-orbital (3e) CASSCF. 

\subsubsection{Dynamic correlation (NEVPT2)}

The SA-CASSCF(4e,3o) energy of the electronic states was corrected by dynamic correlation using 2nd order N-electron valence perturbation theory (NEVPT2). Among the different NEVPT2 implementations~\cite{Angeli-2001a}, we applied the so-called strongly contracted (SC) NEVPT2 method. 

\subsubsection{Choice of the basis set}

The CASSCF-NEVPT2 results, especially the relative energy levels of electronic states, are known to be basis set dependent. Nevertheless, in our previous CASSCF-NEVPT2 study on the \ce{NV-} defect in diamond~\cite{benedekAccurateConvergentEnergetics2024}, we found that the use of the moderate basis set size of cc-pVDZ~\cite{cc-pVDZ} is reasonable, as it practically recovers the basis set limit at cluster sizes comparable to MODEL-2. Therefore, we also sticked to the use of cc-pVDZ in this work.

\subsubsection{Additional technical details}

If density fitting approximations are applied, which considerably accelerate the computation presented in this work at a negligible loss in accuracy, it is also mandatory to choose appropriate auxiliary basis sets. In our calculations, the construction of Coulomb- and exchange integrals during self-consistent field (SCF) procedures was carried out in the RIJCOSX~\cite{NEESE2009} framework, with def2/J~\cite{Weigend2006} auxiliary basis set. To speed up integral transformations from the atomic to the molecular orbital basis, which can be rate-determining in post-SCF quantum chemical calculations, density fitting was introduced with the cc-pVDZ/C~\cite{Weigend2002} auxiliary basis set.

\subsection{Geometry optimization}

The geometry of the cluster model was optimized for each electronic state separately, at state-specific (SS) CASSCF(4e,3o) level. That is, the orbital optimization of the CASSCF procedure was performed exclusively for one selected electron configuration. $C_s$ symmetry was assumed, at which the $E_x$ and $E_y$ states can be conveniently distinguished by setting two different irreducible representations ($A''$ and $A'$, respectively). We note that the SS-CASSCF energy minimization results in Jahn-Teller distorted $C_s$ symmetrical minima for $E$ states, while we obtain $C_{3v}$ symmetrical geometries for $A$ states, in line with group theory considerations. The coordinates of the optimized geometries of the clusters at each electronic state can be found in section S7.

\subsection{Theory of zero-field splitting parameters}

The calculation of zero-field splitting parameters (D tensor, E tensor) requires spin-orbit coupled (SOC) and spin-spin coupled (SSC) states and their energy levels. 
In wavefuncion theory, the SOC and SSC effects are introduced in a post-process manner, within the framework of the quasi-degenerate perturbation theory (QDPT)~\cite{Roemelt_2013}. The QDPT treatment assembles a SOC + SSC matrix from the non-relativistic CASSCF eigenstates $\left({\Psi_I}^{SM}\right)$ as
\begin{equation}
\mel{\Psi_I^{SM}}{\hat{H}_{CASSCF}+\hat{H}_{SOC}+\hat{H}_{SSC}}{\Psi_J^{S'M'}} = \\ \delta_{IJ}\delta_{SS'}\delta_{MM'}E_I^S + \\ \mel{\Psi_I^{SM}}{\hat{H}_{SOC}+\hat{H}_{SSC}}{\Psi_J^{S'M'}}  \\ 
\end{equation}
where the indices $I/J$, $S/S'$, and $M/M'$ refer to the number of the CASSCF root, its spin state, and spin sublevel, respectively. Even though the spin sublevel dependence does not appear explicitly in non-relativistic CASSCF wavefunctions, it can be introduced by means of the Clebsch-Gordan coefficients, in accordance with the Wigner-Eckart theorem~\cite{doi:Neese1998}.  In our study, the $\langle \Psi_I^{SM}|{\hat{H}_{SOC}+\hat{H}_{SSC}}|\Psi_J^{S'M'}\rangle$ matrix elements are calculated between the previously obtained CASSCF roots using the spin-orbit mean-field (SOMF) approximation~\cite{Neese_2005}. Finally, $E_I^S$ in the above equation refers to the (non-relativistic) electronic energy at CASSCF-NEVPT2 level.

The diagonalization of the QDPT matrix yields the coupled states as eigenvectors and the respective energy levels as eigenvalues, from which the ZFS parameters can be extracted as energy differences. Importantly, since CASSCF implementations work in the restricted open-shell formalism (that is, the orbitals of $\alpha$ and $\beta$ spin channels are equal), the spin contamination issues of ZFS calculations, commonly observed for DFT~\cite{biktagirov_spin_2020}, are avoided by construction.  

\subsection{Theory of transition rates between electronic states}

The relaxation pathways, initiated by the photonic excitation of the \ce{ClV+} defect, involve the combination of two different processes: photoluminescence (PL; radiative relaxation towards a lower-energy species of the same multiplicity), internal conversion (IC; nonradiative relaxation towards a lower-energy species of the same multiplicity) and intersystem crossing (ISC; nonradiative relaxation towards a different multiplicity). Fermi's golden rule can be applied to all three transitions~\cite{Schatz1993-kb}, according to which the rate (k) of transition, which occurs as a jump between two energetically close phonon levels (n,m) of the initial (i) and final (f) electronic states, is proportional to i) the overlap of the phonon wavefunctions ($\psi_n$,$\psi_m$), ii) the strength of coupling between the electronic wavefunctions ($\Psi_i$,$\Psi_f$). 
\begin{equation}
k_{in\rightarrow fm} = \frac{2\pi}{\hbar}\left|\langle \Psi_i \psi_n |\hat{H_{c}}| \Psi_f \psi_m \rangle\right|^2 \delta(E_{in}-E_{fm})    
\end{equation}
In the expression above, $\hat{H}_{c}$ represents the Hamiltonian that couples the initial electron-phonon product wavefunction ($in$) to the final state ($fm$), while the Dirac delta of $\delta(E_{in}-E_{fm})$ stands for the energy conservation law. In particular, $\hat{H_{c}}$ corresponds to transition dipole moment, nonadiabatic coupling and spin-orbit coupling in the case of PL, IC and ISC, respectively. 

Taking into account that the transition process involves all possible phonon state transitions, the full rate arises as
\begin{equation}
k_{i\rightarrow f} = \sum_{n,m}p_n(k_{in\rightarrow fm})    
\end{equation}
where $p_n$ is the temperature-dependent Boltzmann factor of the $n$th phonon level. However, at low temperatures, the sum over $n$ can be neglected and it is a reasonable assumption that practically all transitions occur from the zero-point vibrational level of $i$. 

In this work, we apply the 1D effective phonon approximation~\cite{PhysRevB.104.045303}, where the transition occurs on the reaction coordinate connecting the equilibrium geometries of the initial and final state. Consequently both $i$ and $f$ possess only one vibrational mode, along which the coupling was assumed to change linearly (Herzberg-Teller formalism). An advantage of this approach is that computation of rate constants requires only two single-point calculations, i.e. one at the initial and one at the final geometry. The formulation of ISC, IC and PL rates under these constraints~\cite{Smart2021,PhysRevB.100.081407,PhysRevB.90.075202,PhysRevB.110.184302} is detailed below. 

We note here that dynamic Jahn-Teller systems (${}^3\bar{E}, {}^1\bar{E}$) require additional treatment as they continuously rotate around the high-symmetry point via their $\bar{A}'$ and $\bar{A}''$ states and therefore do not possess a well-defined equilibrium geometry; in these cases, we calculated all conceivable single-phonon routes involving the previously optimized Jahn-Teller distorted geometries, and searched for the maximal possible rate.

\subsubsection{Intersystem crossing}

The rate of intersystem crossing, 
corresponding to the relaxation of an excited state towards a state of another multiplicity (triplet-to-singlet conversion, or vice versa), can be calculated based on Fermi's golden rule as
\begin{equation}
k_{{\rm ISC},i \rightarrow f} = \sum_{m}\frac{2 \pi}{\hbar} \delta(E_{i0}-E_{fm}) \\ {\left|\int_{Q=-\infty}^{Q=+\infty} \psi_{i0}(Q) \psi_{fm}(Q) \lambda_{if} (Q,m_S) dQ \right|^2}   
\label{eq:ISC}
\end{equation}
where $E_{i0}$ and $E_{fm}$ refer to the energy level of the zero-point vibrational level of the initial state of emission, and the energy of the $m$ the vibrational level of the final state of emission, respectively. $\psi$ denotes the vibrational wavefunction, which depends on the reaction coordinate (Q), and can be analytically calculated assuming the harmonic movement of the nuclei. The spin-orbit coupling matrix element, ${\lambda_{if}}$, which is to be determined between the electronic wavefunctions of excited and ground states, characterizes the probability of transition. In general, this is also a reaction coordinate dependent parameter. Furthermore, $\lambda$ also depends on the spin sublevel ($m_S$) of the triplet state: $m_S=\pm 1$ and $m_S = 0$ behave as two competing channels of ISC, the relative rate of which gives rise to the spin polarization cycle and the ODMR sign. Accordingly, the matrix elements are often distinguished in the literature as $\lambda_\perp$ and $\lambda_\parallel$ for $\pm 1$ and $0$ sublevels, respectively.   

For practical considerations, the $\delta$ function, which corresponds to a Dirac delta in Fermi's original formulation, was replaced by a Gaussian function: 

\begin{equation}
\delta(E_{in}-E_{fm})\approx(\sqrt{2\pi}\eta)^{-1}e^{-\frac{(E_{in}-E_{fm})^2}{2\eta^2}}
\end{equation}
\noindent
Herein, the square root of variance is represented by a "linewidth" ($\eta$), which we set to be equal to the effective phonon energy of the final state, in accordance with the general recommendations for Fermi's golden rule~\cite{Illg2016-bg}.

\subsubsection{Photoluminescence}

Fermi's golden rule for photoluminescence is somewhat analogous to that for ISC, but additional terms appear due to the appearance of the photon in the final state
\begin{equation}
k(\hbar\omega)_{{\rm PL},i\rightarrow f} = \sum_{m}\frac{1}{4\pi \epsilon_0}\frac{4\omega^3n}{3\hbar c^3} \delta(E_{i0}-(E_{fm}+\hbar\omega)) \\ \left|{\int_{Q=-\infty}^{Q=+\infty} \psi_{i0}(Q) \psi_{fm}(Q) \mu_{if} (Q) dQ}\right|^2   
\label{eq:PL1}
\end{equation}
where $\hbar\omega$ refers to the energy of the emitted photon, and ${\mu_{if}}$ is the electric transition dipole moment, which determines the probability of transition in the case of PL. Among the constants in the prefactor, $n$ stands for the refractive index of the environment (for reference, $n=2.56$ applies to the 4H-SiC lattice at $\approx$ 1550 nm wavelength), while $\epsilon_0$ and $c$ are familiar universal constants (vacuum permittivity, speed of light).

The above phonon energy dependent partial rate (the dimension of which is $\frac{1}{energy \times time}$) is to be integrated over the full spectrum in the last step of the PL calculation. Thus, the overall rate of photoluminescence ($k_{PL}$) is

\begin{equation}
k_{{\rm PL}, i \rightarrow f} = \int_{0}^{\infty} k(\hbar\omega)_{{\rm PL}, i \rightarrow f}d(\hbar\omega)
\label{eq:PL3}
\end{equation}

\subsubsection{Internal conversion}

The rate of internal conversion is determined by nonadiabatic coupling matrix elements, but Fermi's golden rule can also be rewritten based on the overlap (S) of electronic wavefunctions ($\Psi$):

\begin{equation}
S_{if}(Q)=\bra{\Psi_i(Q=0)} \ket{\Psi_f(Q)}   
\end{equation}

In the expression above, $Q=0$ corresponds to a reference nuclear configuration, which is typically chosen as the initial or final equilibrium geometry. Using $S_{if}$ and the electronic energy change of the IC process ($\Delta E_{if}$), the rate can be expressed as~\cite{RevModPhys.73.515,PhysRevB.90.075202}  

\begin{equation}
k_{{\rm IC},i \rightarrow f} = \sum_{m}\frac{2 \pi}{\hbar} \Delta E_{if}^2\delta(E_{i0}-E_{fm})  \\ { \left| \int_{Q=-\infty}^{Q=+\infty} \psi_{i0}(Q) \psi_{fm}(Q) S_{if} (Q) dQ\right|^2}   
\label{eq:IC}
\end{equation}

We note that the overlap was estimated based on the coefficients of individual determinants in the CASSCF wavefunction. Both the initial and the final wavefunction was requested in canonical orbital basis, for the sake of compatibility across different geometries. 

\subsubsection{Software}
The self-developed scripts for evaluation of harmonic oscillator overlaps, integration and evaluation of Fermi rule rates are made available at \url{\codeurl} for the interest of the reader.

\subsection{Sample input files (ORCA 6.0.1.)}
\label{SM:sect:input}

\subsubsection{Geometry optimization (SS-CASSCF) with 3 frontier orbitals in active space}

! cc-pVDZ moread def2/J usesym opt \\
\%moinp ''Initial-DFT-orbitals.gbw'' \\
\%casscf \\
nel 4 \\
norb 3 \\
mult 1 \quad \quad \quad  \#1 for singlet, 3 for triplet states \\
irrep 0 \quad \quad \quad  \#0 for $A'$, 1 for $A''$ states \\
nroots 2 \quad \quad \quad  \#Which state of the specified irrep is optimized (1: lowest-energy state; 2: second lowest state; etc.)  
weights[0]=0,1 \quad \quad \quad  \#Setting the weight of the state of interest to 100\%. \\
maxiter 150 \\ 
printwf det \\
end \\
\%geom \\
Constraints \{C \emph{number\_of\_fixed\_atoms} C\} end \quad \quad  \#Fixed atoms: hydrogens and heavy atoms of the outermost shell\\
end \\
\%sym \\
pointgroup ''cs'' \\ 
end \\
* xyz 1 3 [Symmetrized geometry guess (xyz format)] * \\

\subsubsection{Single-point CASSCF-NEVPT2 calculation with 3+3 electronic states}

\# Notes: The input below also requests transition matrix elements that can be used for transition rate calculations among different electronic states. "Dotrans all" and "dosoc true" refer to transition dipole moment and SOC matrix element computation, respectively. "Dosoc true" and "dossc true" introduces spin-orbit and spin-spin coupling, respectively, based on which zero-field splitting can be determined. The below run without "ptmethod sc\_nevpt2" produces SA-CASSCF results without dynamic correlation (NEVPT2).\\
\\
! rijcosx cc-pVDZ def2/J moread cc-pVDZ/C \\
\%moinp ''CASSCF\_orbitals\_of\_optimized\_geometry.gbw'' \\
\%casscf \\
nel 4 \\
norb 3 \\
mult 3,1 \\
nroots 3,3 \\
ptmethod sc\_nevpt2 \\
actorbs canonorbs \\
printwf det \\
dotrans all \\
rel \\
dosoc true \\
dossc true \\
printlevel 4 \\
end \\
nevpt \\
d3tpre 1e-14 \\
d4tpre 1e-14 \\
end \\
end \\
* xyz 1 3 [Optimized geometry (xyz format)] * \\

\newpage

\subsection{Effect of defect configuration ($hh$, $kk$, $kh$, $hk$)}
\label{SM:sect:k_and_h}

\begin{table*}[h!]
\setlength\extrarowheight{1pt}
\centering
\begin{tabular}{|l|c|c|c|c|}
\hline
 Property &  MODEL-1 ($kk$) & MODEL-1 ($hh$) & MODEL-1 ($kh$) & MODEL-1 ($hk$)   \\
\hline
$^3\bar{A}_2$ vertical energy [eV] & 0.00 & 0.00 & 0.00 & 0.00 \\
$^1\bar{E}$ vertical energy [eV] & 0.38 & 0.39 & 0.36,0.37 & 0.37,0.40 \\
$^1\bar{A}_1$ vertical energy & 1.26 & 1.32 & 1.18 & 1.28\\
$^3\bar{E}$ vertical energy & 1.27 & 1.33 & 1.21,1.22 & 1.25,1.33 \\
\hline
$^3\bar{A}_2$ D tensor [GHz] & 1.39 & 1.45 & 1.47 & 1.45 \\
$^3\bar{A}_2$ E tensor [MHz] & 0.0 & 0.0 & 24.6 & 27.6 \\
\hline
\end{tabular}
\caption{Demonstration on the robustness of key defect characteristics (vertical energy levels and zero-field splitting parameters) against the defect configuration ($kk$, $hh$, $kh$ or $hk$). The cluster model of MODEL-1 size was used for the tests. All numbers were calculated at [3S+3T]-CASSCF(4e,3o)-NEVPT2 level of theory.}
\label{SM:table:configurations}  
\end{table*}

\newpage

\section{Supplement to section III.B.: Calculation of relative energy levels including relaxation effects}
\label{SM:sect:casscf-nevpt2-zpe}

\subsection{Characteristic geometry parameters: C-C distances next to the vacancy}

\begin{table*}[h]
\setlength\extrarowheight{1pt}
\centering
\begin{tabular}{|l|c|c|}
\hline
 Relaxed state &  MODEL-1 & MODEL-2   \\
\hline
$^3\bar{A}_2$ & 3.26 (x3) & 3.28 (x3) \\
$^3\bar{E}\rightarrow{}^3\bar{A}''$ & 3.39 (x2), 3.34 &  3.44 (x2), 3.37\\
$^3\bar{E}\rightarrow{}^3\bar{A}'$  & 3.38 (x3)* & 3.43 (x3)* \\
\hline
$^1\bar{E}\rightarrow{}^1\bar{A}''$  & 3.34 (x2), 3.21 & 3.38 (x2), 3.24 \\
$^1\bar{E}\rightarrow{}^1\bar{A}'$ & 3.30 (x2), 3.40 & 3.35 (x2), 3.42 \\
$^1\bar{A}_1$  & 3.38 (x3) & 3.44 (x3) \\
\hline
\end{tabular}
\caption{Carbon-carbon interatomic distances (\AA) in the innermost shell of atoms in state-specifically optimized geometries (SS-CASSCF(4e,3o)/cc-pVDZ), for the $kk$ configuration. * These structures are not $C_{3v}$ symmetrical; however, interestingly, the Jahn-Teller distortion mainly manifests in the asymmetry of the Si-C bond lengths in the vicinity of the vacancy.}
\label{SM:table:distances}  
\end{table*}

\newpage
\subsection{Calculation of relaxed excitation energies}

The energy levels relative to the ground state ($(1){}^3\bar{A}_2$) presented in Fig. 2 of the main text were determined as follows. 

In the first step, the energy levels at $(1){}^3\bar{A}_2$ geometry were calculated, using NEVPT2 on top of a state-averaged CASSCF(4e,3o) wavefunction (3 singlet, 3 triplet roots). 

Secondly, relaxation within $C_{3v}$ symmetry was calculated at state-specific CASSCF level. While the SS-CASSCF equilibrium geometry of $\bar{A}$ states has $C_{3v}$ symmetry by construction, additional geometry optimization is required for $\bar{E}$ states to gain a $C_{3v}$ symmetrical energy minimum. In the case of $(1){}^3\bar{E}$ and $(1){}^1\bar{E}$, we managed to locate the minimum-energy crossing point (MECP) of $\bar{E}_x$ and $\bar{E}_y$ by geometry optimization with 50-50\% weighing, and the relaxation within $C_{3v}$ symmetry constraint was studied on these structures. 

Thirdly, the energy contribution of Jahn-Teller distortion of $E$ states was calculated, which is given by the energy difference between $A'$/$A''$ and MECP geometries at SS-CASSCF level.

Tab. \ref{SM:table:calc3} summarizes the results of relaxation-corrected CASSCF-NEVPT2 calculations, which altogether describe the relative energy levels comparably to experimental ZPLs.

\begin{table*}[h]
\setlength\extrarowheight{1pt}
\centering
\resizebox{\textwidth}{!}{%
\begin{tabular}{|c|c|c|c|c|c|c|c|c|}
\hline
State &  \multicolumn{4}{|c|}{MODEL-1} & \multicolumn{4}{|c|}{MODEL-2}   \\
\cline{2-9}
& $E_V$ & $E_{\rm rel}$($C_{3v}$) & $E_{\rm rel}$(JT) & $E_{\rm tot}$ & $E_V$ & $E_{\rm rel}$($C_{3v}$) & $E_{\rm rel}$(JT) & $E_{\rm tot}$   \\
\hline
$^3\bar{A}_2$ & $0.00$ & -& - & $0.00$ & $0.00$&-&-&$0.00$  \\
\hline
${}^1\bar{E}_x\rightarrow{}^1\bar{A}''$ & $0.38$ &$0.00$&$-0.10$&$0.28$&$0.33$&$-0.01$&$-0.12$&$0.20$ \\
\hline
${}^1\bar{E}_y\rightarrow{}^1\bar{A}'\phantom{'}$ & $0.38$&$0.00$&$-0.09$&$0.29$&$0.33$&$-0.01$&$-0.14$&$0.18$ \\
\hline
${}^1\bar{A}_1$ & $1.26$&$-0.07$&-&$1.19$&$1.07$&$-0.10$&-&$0.97$ \\
\hline
${}^3\bar{E}_x\rightarrow{}^3\bar{A}''$ & $1.27$&$-0.12$&$-0.04$&$1.11$&$1.11$&$-0.17$&$-0.04$&$0.94$ \\
\hline
${}^3\bar{E}_y\rightarrow{}^3\bar{A}'\phantom{'}$ &$1.27$ &$-0.12$&$-0.08$&$1.07$&$1.11$&$-0.17$&$-0.09$&$0.85$ \\
\hline

\hline

\end{tabular}}
\caption{Energy contributions to the relaxed electronic spectrum presented in Fig. 3 of the main text. $E_V$: vertical energy ($(1){}^3\bar{A}_2$ geometry, SA-CASSCF-NEVPT2 level of theory). $E_{\rm rel}$($C_{3v}$): energy decrease due to geometry relaxation within $C_{3v}$ symmetry (SS-CASSCF level of theory). $E_{\rm rel}$(JT): energy decrease due to Jahn-Teller geometry distortion (SS-CASSCF level of theory). $E_{\rm tot}$: total energy relative to $(1)^3\bar{A}_2$, calculated as the the sum of $E_V$, $E_{\rm rel}$($C_{3v}$) and $E_{\rm rel}$(JT). All presented values are given in eV.}
\label{SM:table:calc3}  
\end{table*}

\newpage

\section{Calculation of transition rates}
\label{SM:sect:rates}

\begin{table*}[h]
\caption{Details of the computation of transition rates of \textbf{\ce{ClV+} in 4H-SiC} (PL: photoluminescence; ISC: intersystem crossing) based on the 1D effective phonon approximation. $Q$: Sqare-root-mass-weighted displacement of atomic coordinates between initial ($i$) and final ($f$) states. $h\omega$: Energy of the fitted effective (harmonic) phonon, connecting initial and final equilibrium geometries. $\Delta E$: Energy shift  between the minima of initial and final potential energy surfaces. TME: Transition matrix element, which stands for transition dipole moment (in the case of PL), spin-orbit coupling (in the case of ISC), or wavefunction overlap (in the case of IC). $k$: Transition rate (frequency). $\tau$: Half-life of initial state. Spin-forbidden transitions (where the transition matrix element is zero at both geometries) are omitted. Note: Routes connecting $A'$ and $A''$ states assume $C_s$ symmetry at all internal geometries. For other pathways of $A' \leftrightarrow A"$, which result in the reorientation of the mirror plane, the parabolic approximation of the potential energy surfaces is not applicable. (In practice, we set the orientation of $A'$ and $A''$ to the same mirror plane before applying the 1D Fermi's golden rule.)      }
\setlength\extrarowheight{1pt}
\centering
\resizebox{\textwidth}{!}{%
\begin{tabular}{|c|c|c|c|c|c|c|c|c|c|c|c|}
\hline
Transition & Pathway & Type & $m_S$ & \multicolumn{8}{|c|}{MODEL-2}  \\
\cline{5-12}
&&&& $Q$ [$\sqrt{\rm amu}$\AA] & $\hbar\omega_i$ [meV] & $\hbar\omega_f$ [meV] & $\Delta E$ [eV] & TME($i$) & TME($f$) & $k$ & $\tau$  \\
\hline
\textbf{${}^3\bar{E} \rightarrow {}^3\bar{A}_2$} & ${}^3\bar{A}'' \rightarrow {}^3\bar{A}_2$ & PL &$0,\pm1$ &0.973&42.6&37.0&0.90&7.0 D& 6.6 D& \textbf{33.1 MHz}& \textbf{30.2 ns}\\
&${}^1\bar{A}' \rightarrow {}^3\bar{A}_2$ & PL & $0,\pm1$& 1.087 & 42.9&38.6&0.85&7.5 D&6.6 D&22.3 MHz&44.8 ns  \\
\cline{2-12}
 & ${}^3\bar{A}'' \rightarrow {}^3\bar{A}_2$ & IC & $0,\pm1$ &0.973&42.6&37.0&0.90&0.0475&0.0000&\textbf{7.07e-2 Hz}&\textbf{14.1 s}  \\
\hline
\textbf{${}^3\bar{E} \rightarrow {}^1\bar{A}_1$}&${}^3\bar{A}'' \rightarrow {}^1\bar{A}_1$ & ISC & $\pm1$ &0.353&50.0&50.3&-0.07&19.0 GHz&17.2 GHz&\textbf{61.1 MHz}&\textbf{16.4 ns}  \\
&${}^3\bar{A}' \rightarrow {}^1\bar{A}_1$ & ISC & $\pm1$ &0.400&69.5&88.5&-0.12&26.4 GHz&17.2 GHz&24.7 MHz&40.5 ns  \\
\cline{2-12}
&${}^3\bar{A}'' \rightarrow {}^1\bar{A}_1$ & ISC & $0$ &0.353&50.0&50.3&-0.07&1.1 GHz&0.0 GHz&\textbf{51.8 kHz}&\textbf{19.3 $\mu$s}  \\
\hline
\textbf{${}^3\bar{E} \rightarrow {}^1\bar{E}$}&${}^3\bar{A}'' \rightarrow {}^1\bar{A}''$ & ISC & $\pm1$ &0.827&74.1&48.6&0.70&45.5 GHz&45.3 GHz&556 Hz&1.8 ms  \\
&${}^3\bar{A}'' \rightarrow {}^1\bar{A}'$ & ISC & $\pm1$ &0.909&49.8&48.3&0.72&44.1 GHz&43.7 GHz&192 kHz&5.2 $\mu$s  \\
&${}^3\bar{A}' \rightarrow {}^1\bar{A}''$ & ISC & $\pm1$ &0.918&41.9&36.5&0.65&46.7 GHz&42.9 GHz&0.37 Hz&2.7 s  \\
&${}^3\bar{A}' \rightarrow {}^1\bar{A}'$ & ISC & $\pm1$ &1.252&64.4&46.0&0.67&40.2 GHz&45.6 GHz&\textbf{43.2 MHz}&\textbf{23.2 ns}  \\
\cline{2-12}

&${}^3\bar{A}'' \rightarrow {}^1\bar{A}'$ & ISC & $0$ &0.909&49.8&48.3&0.72&23.8 GHz&29.8 GHz&\textbf{138 kHz}&\textbf{7.3 $\mu$s}  \\
&${}^3\bar{A}' \rightarrow {}^1\bar{A}''$ & ISC & $0$ &0.918&41.9&36.5&0.65&25.7 GHz&25.9 GHz&0.27 Hz&3.7 s  \\

\hline
\textbf{${}^1\bar{A}_1 \rightarrow {}^1\bar{E}$}&${}^1\bar{A}_1 \rightarrow {}^1\bar{A}''$ & PL &$0,\pm1$ &0.997&46.7&41.5&0.77&6.2 D& 4.7 D& \textbf{7.60 MHz}& \textbf{132 ns}\\
&${}^1\bar{A}_1 \rightarrow {}^1\bar{A}'$ & PL & $0,\pm1$&0.983&67.5&46.2&0.79&6.2 D& 3.6 D& 4.76 MHz&210 ns  \\
\cline{2-12}

&${}^1\bar{A}_1 \rightarrow {}^1\bar{A}'$ & IC & $0,\pm1$&0.983&67.5&46.2&0.79&0.000&0.377&\textbf{601 MHz}&\textbf{1.66 ns}  \\
\hline

\cline{2-12}
${}^1\bar{A}_1 \rightarrow {}^3\bar{A}_2$&${}^1\bar{A}_1 \rightarrow {}^3\bar{A}_2$ & ISC & $0$ &1.056&35.3&35.6&0.97&66.9 GHz&67.3 GHz&\textbf{0.061 Hz}&\textbf{16.5 s}  \\
\hline
\textbf{${}^1\bar{E} \rightarrow {}^3\bar{A}_2$}&${}^1\bar{A}'' \rightarrow {}^3\bar{A}_2$ & ISC & $\pm1$ &0.672&49.2&38.9&0.20&17.1 GHz&24.0 GHz&108 MHz& 9.3 ns \\
&${}^1\bar{A}' \rightarrow {}^3\bar{A}_2$ & ISC & $\pm1$ &0.941&38.0&41.5&0.18&10.0 GHz&24.0 GHz&\textbf{381 MHz}&\textbf{2.6 ns}   \\
\cline{2-12}

&${}^1\bar{A}' \rightarrow {}^3\bar{A}_2$ & ISC & $0$ &0.672&49.2&38.9&0.20&8.1 GHz&0.0 GHz&\textbf{364 kHz}&\textbf{2.7 $\mu$s}  \\
\hline
\end{tabular}}
\label{SM:table:Rates_1}  
\end{table*}


\newpage

\section{Calculation of photoluminescence characteristics}
\label{SM:sect:PL}

The characteristic parameters of the triplet and singlet photoemission of \ce{ClV+}@4H-SiC are summarized in Tab.~\ref{SM:tab:PL_parameters}.  The HR and DW factors can be obtained from the data of Tab.~\ref{SM:table:Rates_1} as

\begin{equation}
{\rm HR} = \frac{\omega_f Q^2}{2\hbar} 
\end{equation}

and

\begin{equation}
{\rm DW} = e^{-\rm HR}
\end{equation}

The uncertainty range indicated in the Table derives from the multiple possible Jahn-Teller distorted geometries of $E$ states.

\begin{table*}[h]
\setlength\extrarowheight{0pt}
\centering
\begin{tabular}{|l|c|c|}
\hline
Parameter [unit] & ${}^3\bar{E}\rightarrow{}^3\bar{A}_2$ & ${}^1\bar{A}_1\rightarrow{}^1\bar{E}$ \\
\hline
ZPL [eV] & 0.85-0.90 & 0.77-0.79 \\
PSB$_{\rm max}$ [eV] & 0.64-0.75 & 0.56-0.57 \\
ZPL $-$ PSB$_{\rm max}$ [eV] & 0.15-0.21 & 0.20-0.23 \\
Q [$\sqrt{\rm amu}$\AA] & 0.97-1.09&0.94-1.00 \\
$\hbar\omega_{\rm f}$ [meV] &37-39&41-46 \\
HR factor [-] & 4.2-5.4 & 4.9-5.0\\
DW factor [-] & 0.5-1.5\% & 0.7-0.8\% \\
\hline
\end{tabular}
\caption{Photoluminescence parameters of the \ce{ClV+} defect, as obtained at CASSCF-NEVPT2/cc-pVDZ level for different model sizes.}
\label{SM:tab:PL_parameters}  
\end{table*}

\newpage

\section{Corrected DFT Zero-field splitting parameters for all defect configurations}

\subsection{DFT computational details}

Supporting DFT calculations were performed starting from optimized geometries from Ref.~\cite{Oscar}, consisting of 576 atom 4H-SiC supercells. Calculations using the PBE functional~\cite{PBE} were performed on cells with dimensions of 18.56 \AA\ along $ab$ lattice directions and 20.25 \AA\ along the $c$-direction.
DFT calculations in VASP were performed with a 420 eV energy cutoff for the plane-wave basis set, with electron density and forces optimized with thresholds of $10^{-6}$ eV and $0.01$ eV/\AA, respectively. PBE calculations were done on a 2x2x2 $\Gamma$-centered k-point grid.
Excited state calculations were in VASP treated with the constrained-occupation technique, strictly (de)occupying the ($a_1$) $e$ orbitals to obtain the excited state of interest.

\subsection{Revised DFT results}

ZFS parameters already reported in Bulancea-Lindvall et al.~\cite{Oscar} likely suffer from spin contamination errors. However, using a spin decontamination technique developed by Biktagirov et al.~\cite{biktagirov_spin_2020} for eliminating such errors in DFT, we also provide corrected DFT values here. We note, however, that the technique requires converging several singlet states using the constrained-occupation approach, and calculating their (non-spin-scaled) ZFS parameters, which can be challenging at the risk of switching the orbital order, as DFT inherently tries to find the lowest-energy state. Therefore, the ZFS parameters for the singlet states, with the exception of the $kk$ configuration, were instead obtained using the frozen wavefunctions in the ${}^3A_2$ ground state. We also note that non-spin-scaled ZFS tensors were calculated via a modified binary, allowing for such calculations even in the singlet-spin case. All such calculations were performed at PBE-level for corresponding PBE-optimized geometries.

Here we provide calculated ZFS parameters as done in VASP with the addition of the spin-decontamination technique~\cite{biktagirov_spin_2020} as a post-processing step. This technique involves eliminating the spin contamination by $D$-tensor substraction by an average of the $D$-tensor of $S-1$ states corresponding to the target spin $S$ state. As seen in this case, the difference in techniques is on the level of 100 MHz, with the fully relaxed case already shown to be highly accurate for similar defects in SiC and diamond\cite{biktagirov_spin_2020}, and errors mostly emerging from the lack of SOC contributions.
\begin{table}[h!]
    \centering
    \begin{tabular}{|l|c|c|c|c|}
    \hline
        Config. & $hh$ & $kk$ & $hk$ & $kh$ \\
        ZFS $D$ (GHz) & * & 1.15 & * & * \\
        ZFS $E$ (MHz) & * & 0 & * & * \\
        $\left({}^3A_2\right)$ ZFS $D$ (GHz) & 1.18 & 1.05 & 1.08 & 1.01 \\
        $\left({}^3A_2\right)$ ZFS $E$ (MHz) & 0 & 0 & 43 & 18 \\
    \hline
    \end{tabular}
    \caption{Spin-decontaminated ZFS parameters for DFT calculated ClV${}^+$ ${}^3A_2$ ground state. Rows denoted by $\left({}^3A_2\right)$ specify calculations not relaxing the density of the constrained singlet states, as opposed to the original instruction by Biktagirov et al.~\cite{biktagirov_spin_2020}. * denotes values not calculated in this study.}
    \label{tab:my_label}
\end{table}

\section{Qualitative ODMR Estimation}
With the corrected DFT ZFS tensors and local hyperfine interactions calculated in Ref.~\cite{Oscar}, a spin model of the ClV${}^+$ electron spin can be constructed and its eigenvalue landscape can be examined for possible ODMR transitions.

\begin{table}[h!]
    \centering
    \caption{Chlorine substitutional site hyperfine values used in the spin model for the ClV${}^+$, shown here for ${}^{35}$Cl, with the $z$-axis aligned with the $c$-axis and the $x$-axis aligned with the $a$-axis of the host crystal. Values are in MHz. Were obtained in Ref.~\cite{Oscar} using similar DFT settings as in this work, but with the HSE06 functional.}
    \begin{tabular}{|c|cccccc|}
    \hline
        Config. & $A_{xx}$ & $A_{yy}$ & $A_{zz}$ & $A_{xy}$ & $A_{xz}$ & $A_{yz}$ \\
        \hline
        $hh$ & $-16.328$ & $-16.328$ & $-15.752$ & 0 & 0 & 0\\
        $kk$ & $-18.081$ & $-18.082$ & $-16.862$ & 0 & 0 & 0\\
        $hk$ &  $-11.796$ & $-11.994$ & $-12.114$ & $0.172$ & $0.205$ & $0.118$\\
        $kh$ & $-9.292$ & $-8.952$ & $-9.869$ & $0.015$ & $0.001$ & $0.034$\\
        \hline
    \end{tabular}
    \label{tab:Cl-hyperfine}
\end{table}

We assume a spin defect $g$-factor equal to the Landé $g$-factor. As the Cl species in the defect is either ${}^{35}$Cl or ${}^{37}$Cl, both exhibiting an $I=3/2$ nuclear spin, we include this spin into the model as well. These chlorine species differ mainly in the nuclear $g$-factor ($g_{35} = 0.5479162$ and $g_{37} = 0.4560824$) which enters into the nuclear Zeeman contribution and as a scaling factor to the DFT hyperfine tensor obtained from the electron density~\cite{ivadyFirstPrinciplesCalculation2018b}. We therefore note that switching from the more abundant ${}^{35}$Cl (ca.\ 76\%) to ${}^{37}$Cl (ca.\ 24\%), the hyperfine and the size of the Zeeman splitting is expected to decrease by ca.\ 20\%.

Table~\ref{tab:Cl-hyperfine} shows the hyperfine tensors used for ${}^{35}$Cl. Note, however, that the $hk$ and $kh$ tensors are only valid up to $120^\circ$ rotation around the $c$-axis due to their three possible defect axis alignments in the host.

The Hamiltonian of the ClV${}^+$ and Cl spins is constructed as
\begin{equation}\label{eq:spin-hamil}
    H = g_{\rm ClV}\mu_{\rm B}\Vec{S}\cdot \Vec{B} + g_{\rm Cl}\mu_{\rm N} \Vec{I}\cdot \Vec{B} + \Vec{S}D\Vec{S} + \Vec{S}A\Vec{I},
\end{equation}
containing, in order, the electron and nuclear Zeeman contributions, the ZFS and the hyperfine interactions, involving electron $S=1$ spin operators, $\Vec{S} = (S_x, S_y, S_z)$ and nuclear Cl $I=3/2$ spin operators, $\Vec{I} = (I_x, I_y, I_z)$. This Hamiltonian is diagonalized at several magnetic field vectors, and the eigenvalue difference for states $\phi_i$ and $\phi_j$ are weighted by the matrix element of these states for the exchange operator
\begin{equation}
    O_{ij}(\Vec{B}) = \bra{\phi_i(\Vec{B})}S_x \otimes I + I \otimes I_x\ket{\phi_j(\Vec{B})}.
\end{equation}
The matrix element is a qualitative indicator for how the states would be affected by the microwave drive entering as an oscillating magnetic field term in Hamiltonian \eqref{eq:spin-hamil}. Incorporating a Gaussian smearing in the ODMR frequency to describe experimental precision and quantum-enforced uncertainty, a qualitative ODMR estimate $\mathcal{I}$, can be obtained as
\begin{equation}
    \mathcal{I}(\Vec{B},\omega) = \sum_{i>j} \left|O_{ij}(\Vec{B})\right|e^{-(\omega - |E_i(\Vec{B}) - E_j(\Vec{B})|)^2/2\sigma^2},
\end{equation}
with smearing parameter $\sigma$.
We note that a more accurate model can be constructed, incorporating the kinetics of the metastable state, exited state and the rates predicted in this work. However, given the non-quantitative accuracy of the ab initio rates, a qualitative non-dynamic model is deemed sufficient for the extent of this work.

\begin{figure}[h!]
    \centering
    \begin{minipage}{0.48\textwidth}
    \tikz{
    \node [anchor=north west] at (0,0) {\includegraphics[width=\linewidth, clip, trim=1.25cm 0.25cm 1.75cm 1.25cm]{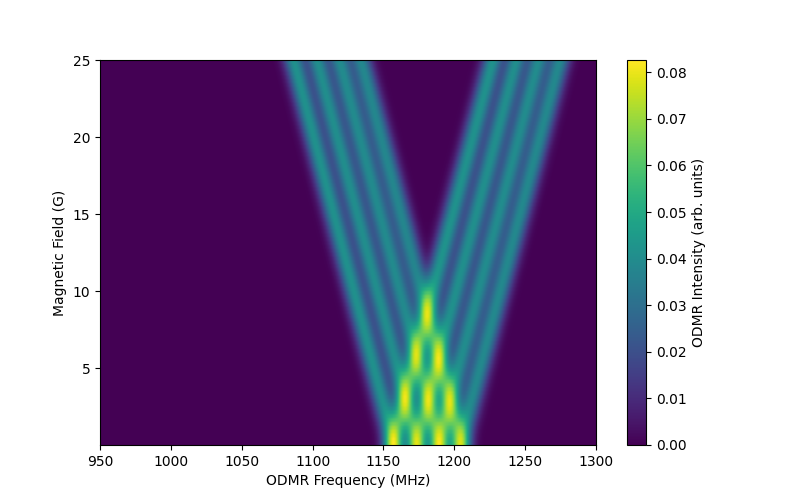}};
    \node [color=white, scale=1.4] at (1.5cm,-1cm) {$hh$};
    }
    \end{minipage}
    \quad
    \begin{minipage}{0.48\textwidth}
        \tikz{
        \node [anchor=north west] {\includegraphics[width=\linewidth, clip, trim=1.25cm 0.25cm 1.75cm 1.25cm]{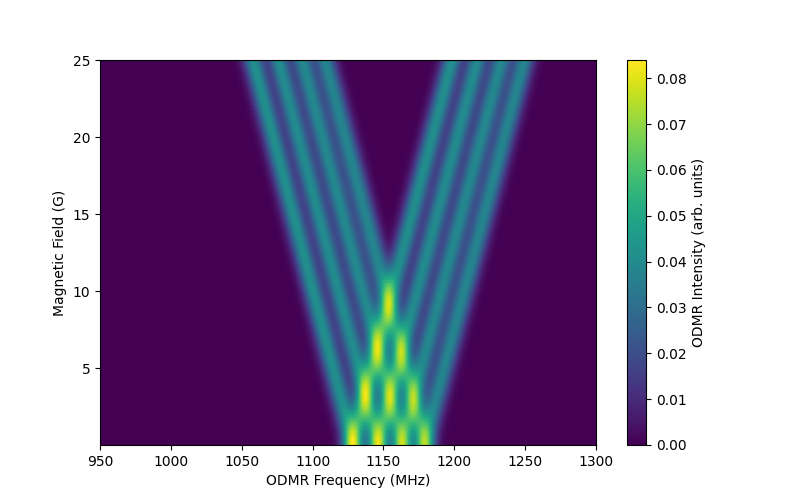}};
        \node [color=white, scale=1.4] at (1.5cm,-1cm) {$kk$};
        }
    \end{minipage}\\
    \begin{minipage}{0.48\textwidth}
    \tikz{
        \node [anchor=north west] {
            \includegraphics[width=\linewidth, clip, trim=1.25cm 0.25cm 1.75cm 1.25cm]{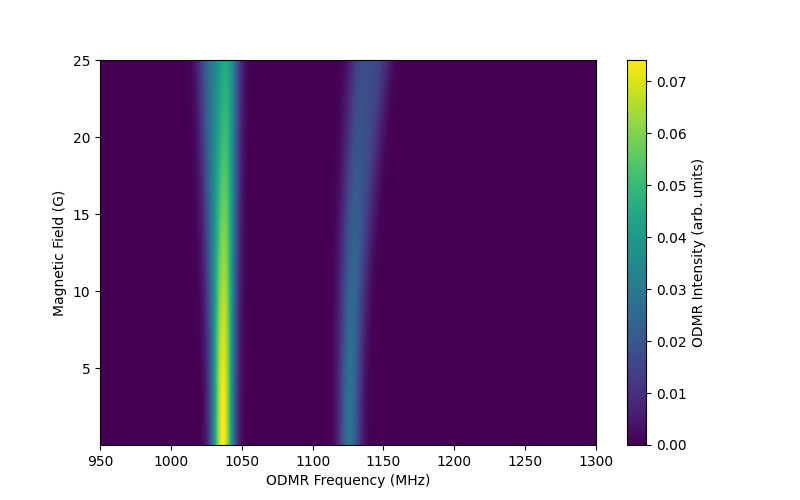}};
    \node [color=white, scale=1.4] at (1.5cm,-1cm) {$hk$};
    }
    \end{minipage}
    \quad
    \begin{minipage}{0.48\textwidth}
    \tikz{
        \node [anchor=north west] {
        \includegraphics[width=\linewidth, clip, trim=1.25cm 0.25cm 1.75cm 1.25cm]{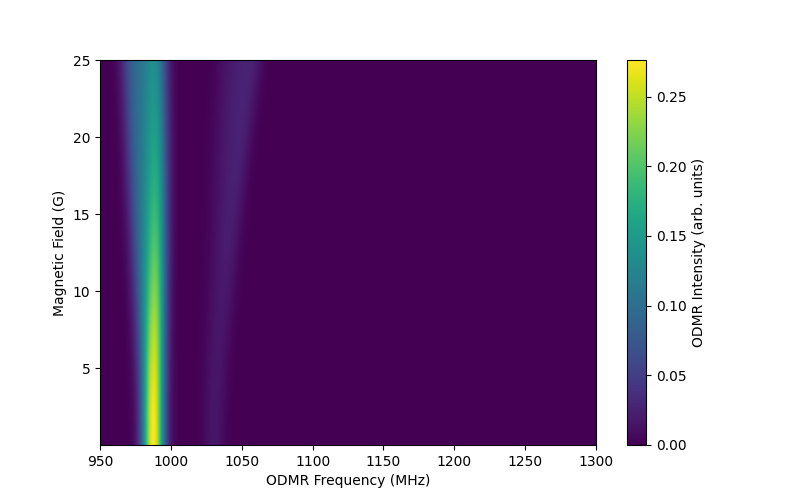}};
        \node [color=white, scale=1.4] at (3.5cm,-1cm) {$kh$};
    }
    \end{minipage}
    \\
    \centering
    \begin{minipage}{0.48\textwidth}
    \tikz{
    \node [anchor=north west] at (0,0) {
        \includegraphics[width=\linewidth, clip, trim=1.25cm 0.25cm 1.75cm 1.25cm]{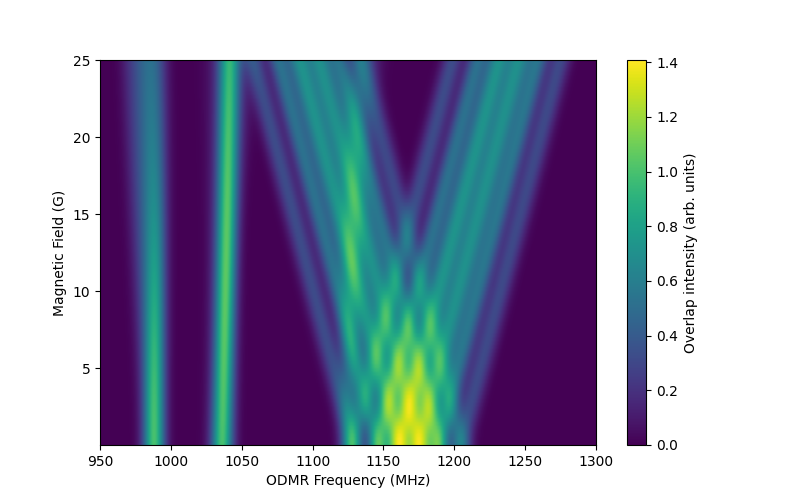}};
    \node [color=white, scale=1.4] at (2.5cm, -1cm) {ClV${}^+$ 5 MHz};
    }
    \end{minipage}
    \quad
    \begin{minipage}{0.48\textwidth}
    \tikz{
    \node [anchor=north west] at (0,0) {
        \includegraphics[width=\linewidth, clip, trim=1.25cm 0.25cm 1.75cm 1.25cm]{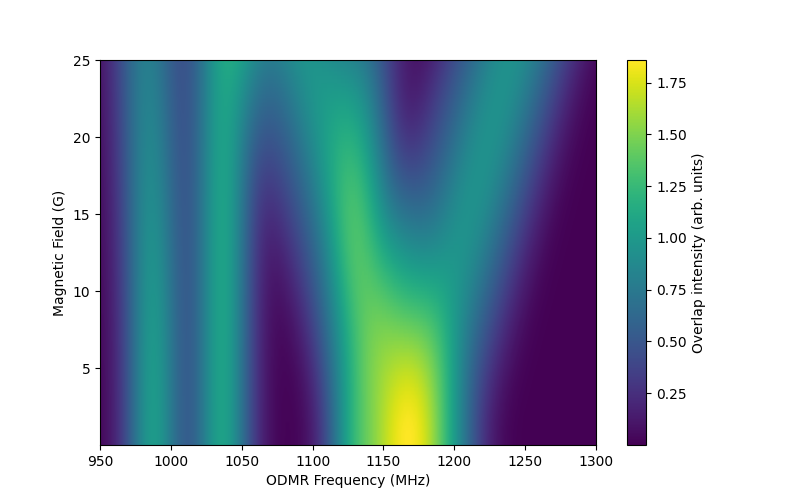}};
    \node [color=white, scale=1.4] at (2.5cm, -1cm) {ClV${}^+$ 15 MHz};
    }
    \end{minipage}
    \caption{Individual and consolidated ODMR spectrum of the ClV${}^+$ configurations, with the magnetic field aligned with the $c$-axis. The bottom figures show combined spectras of all configurations, with 5 MHz and 15 MHz broadening, i.e. at the level where the Cl hyperfine is discernable, and when it is not. Let the reader note that the consolidated plots sum the individual ODMR amplitudes without regarding to differing i) defect concentration ii) configuration brightness iii) or spin contrast between configurations, and can therefore mostly provide a sense of qualitative ODMR characteristics, with limited comparison between the configurations.}
    \label{fig:ClV-odmr}
\end{figure}

With the above described methodology and system parameters, the qualitative reconstruction of low-field ODMR signals of the ClV${}^+$ configurations are displayed in Figure~\ref{fig:ClV-odmr}. For the on-axis configurations, $hh$ and $kk$ the hyperfine of the Cl isotope splits the originally two electron spin $\ket{\pm 1} \leftrightarrow \ket{0}$ ODMR signals into eight lines separated by a multiple of the $A_{zz}$ Cl hyperfine. As these configurations also have a low ZFS $E$-value, emerging only from additional strain or environment electric sources, the eight lines are likely to cross at zero field and create what might appear as a single broad signal depending on the ODMR frequency resolution. Furthermore, the size of the hyperfine-induced splitting, defining the width of the line region if unresolved, may be on the level of the ZFS $D$-value separation between the $hh$ and $kk$ configuration, increasing the difficulty to resolve the two configurations. However, we note that the corrected ZFS still has an error in the range of 100 MHz and may in reality separate enough for the two ODMR configuration signals to be resolved. Alternatively, the dynamics of the system could allow some of these eight lines to be less favored once the system reaches a steady state. For the off-axis configurations, the $E$-value causes a splitting of the line on the same degree or more than the hyperfine-induced splitting, causing a grouping of the eight lines into two regions where the lines are likely unresolved. In addition, at low magnetic fields, the E-value splitting and the orientation of the ZFS axes to the magnetic field applied along the $c$-axis, as plotted in Figure~\ref{fig:ClV-odmr}, the lines will appear relatively constant with respect to the magnetic field and have a significant bias in the allowed transitions. For the $kh$ configuration, the group at a higher ODMR frequency is almost forbidden in relation to the group at lower frequencies, which implies that this configuration may only create one discernible transition line in the ODMR.

\newpage

\bibliography{references}{}

\section{Optimized geometries in xyz format}

\includepdf[pages=-]{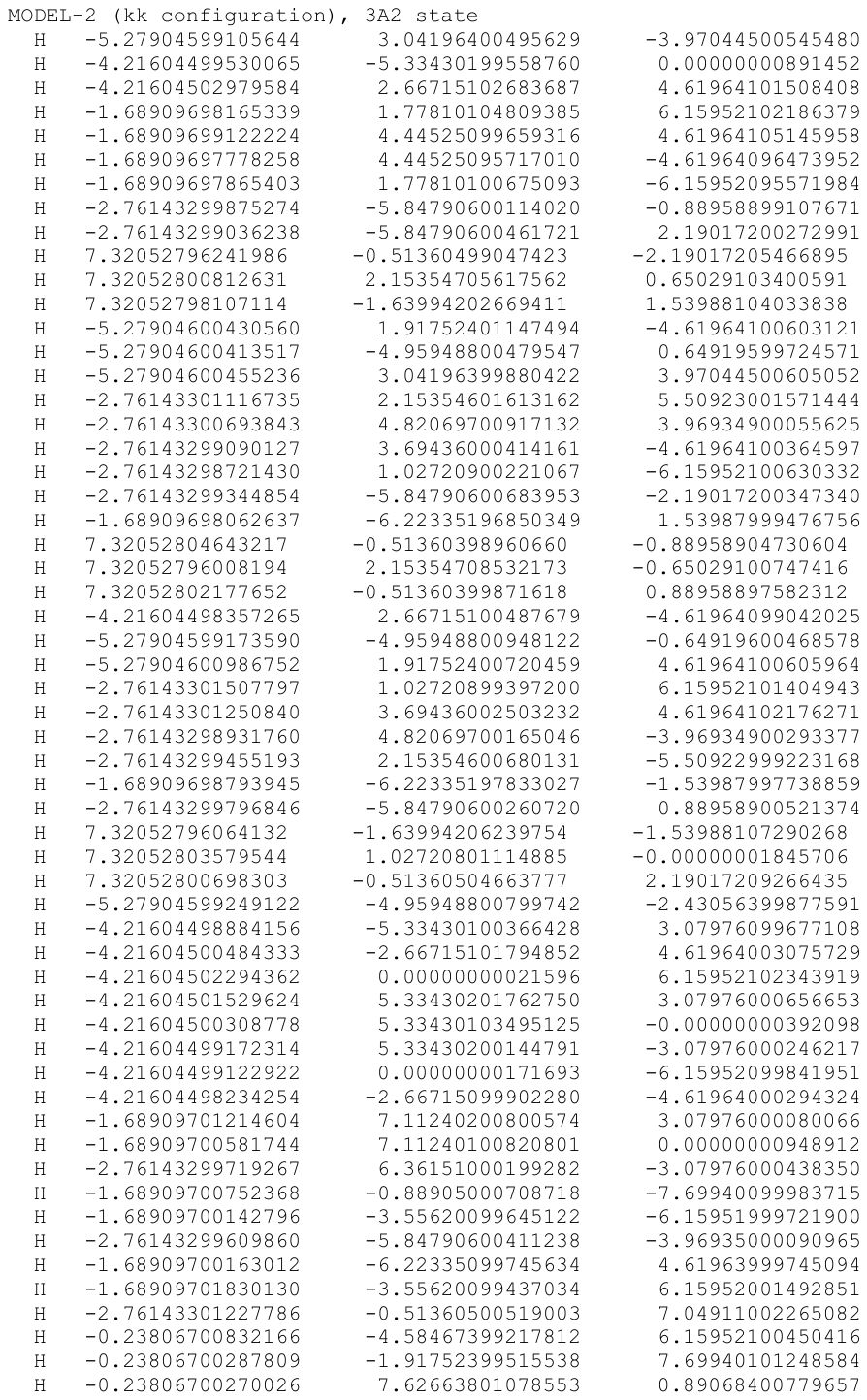}